\documentclass[achemso  ,graphicx, reprint]{revtex4-2}

\usepackage{graphicx}
\usepackage{dcolumn}
\usepackage{bm}

\usepackage[utf8]{inputenc}
\usepackage[T1]{fontenc}
\usepackage{mathptmx}
\usepackage{etoolbox}
\usepackage{color}

\usepackage{amsmath}


\begin{document}


\title{Quantum efficiency and vertical position of quantum emitters in hBN determined by Purcell effect in hybrid metal-dielectric planar photonic structures} 



\author{Domitille G\'erard$^1$, Aur\'elie Pierret$^2$, Helmi Fartas$^1$, Bruno Berini$^1$, St\'ephanie Buil$^1$, Jean-Pierre Hermier$^1$, Aymeric Delteil$^1$}
\affiliation{1. Universit\'e Paris-Saclay, UVSQ, CNRS,  GEMaC, 78000, Versailles, France. \\
2. Laboratoire de Physique de l'\'Ecole Normale Sup\'erieure, ENS, Universit\'e PSL, CNRS, Sorbonne Universit\'e, Universi\'e Paris-Cit\'e, 75005, Paris, France}


\date{\today}

\begin{abstract}
Color centers in hexagonal boron nitride (hBN) advantageously combine excellent photophysical properties with a potential for integration in highly compact devices. Progress towards scalable integration necessitates a high quantum efficiency and an efficient photon collection. In this context, we compare the optical characteristics of individual hBN color centers generated by electron irradiation, in two different electromagnetic environments. We keep track of well-identified emitters that we characterize before and after dry transfer of exfoliated crystals. This comparison provides information about their quantum efficiency -- which we find close to unity -- as well as their vertical position in the crystal with nanometric precision, which we find away from the flake surfaces. Our work suggests hybrid dielectric-metal planar structures as an efficient tool for characterizing quantum emitters in addition to improving the count rate, and can be generalized to other emitters in 2D materials or in planar photonic structures.

\end{abstract}


\maketitle 

\section*{Introduction}

Solid-state single photon emitters (SPEs) are widely seen as pivotal elements for integrated photonic quantum computing~\cite{Aharonovich16,Wang20}. They must however fulfil stringent requirements on their optical properties. This includes a high internal quantum efficiency~$\eta_i$, which is crucial for the scalability of potential applications considering the exponential impact on the success rate of quantum protocols. 

Among solid-state emitters with attractive properties for quantum applications, color centers in hexagonal boron nitride (hBN) are particularly appealing owing to their potential for integration into highly miniaturized devices realized with fabrication techniques that are specific to 2D materials. These SPEs are in most cases bright, stable, and emit up to room temperature~\cite{Tran16, Bourrelier16, Martinez16} -- yet many of them have suboptimal quantum efficiencies~\cite{Tran16, Schell18, Nikolai19, Reimers20}. Of the many classes of SPEs in hBN, the blue-emitting color centers with a ZPL at 436~nm (B-centers) have raised a strong interest since they can be locally generated by electron irradiation~\cite{Shevitski19, Fournier21, Gale22}. They have already proven a high potential for quantum applications, with, among other advantageous characteristics, a narrow wavelength spread~\cite{Fournier21, Horder22}, a high coherence~\cite{Horder22, Fournier22PRB} allowing for single photons with sizable indistinguishability~\cite{Fournier22PRA}, and a potential for top-down integration into photonic structures~\cite{Gerard23, Spencer23}. However, their quantum efficiency has not been inferred to date.

Here, we compare the photoluminescence of well-identified individual B-centers in two different electromagnetic (EM) environments to estimate this figure of merit. Given that the intrinsic nonradiative relaxation is insensitive to the local density of optical states (LDOS), the modifications of the decay rate allows to identify the radiative contribution to the excited state dynamics~\cite{Drexhage70,Tews70, Kunz80, Brokmann04}. This comparison is based on the transfer of hBN crystals hosting SPEs from a dielectric to a metallic substrate realizing a hybrid metal-dielectric planar photonic structure, which strongly modifies the emission properties of the emitters. The observed lifetime modifications are compared to numerical calculations so to infer the quantum efficiency~\cite{Nikolai19}, which we find close to unity. Additionally, the decay acceleration carries information about the emitter vertical position in the flake, which we estimate for multiple emitters with a confidence interval of order of a few nanometers in most cases, as established by maximum likelihood estimation. This allows to establish that stable emitters are found relatively far from the flake surface. Finally, the lossy planar cavity formed by the hybrid metal-dielectric structure allows a count rate enhancement up to a factor~8, depending on the emitter position. 

\section*{Experimental protocol and optical characterization}

The principle behind the present work is to characterize the same SPEs in two different EM environments to extract information from the variations of measured optical properties~\cite{Tran17, Dowran23}. In the absence of any photonic structure (\textit{i.e.} in an homogeneous medium), a quantum emitter can undergo both radiative and non-radiative decay processes. We note $\tau_\mathrm{rad}$ the radiative lifetime, and $\Gamma_\mathrm{rad} = 1/\tau_\mathrm{rad}$. The decay time associated with all intrinsic non-radiative processes, supposed independent of the LDOS, is noted $\tau_\mathrm{nr}$ (with $\Gamma_\mathrm{nr} = 1/\tau_\mathrm{nr}$). In the case of color centers in wide-gap materials, the physical origin of such processes can include intersystem crossing to non-emissive states, as well as coupling to surrounding crystal defects and to interface- or surface-induced processes~\cite{Mohtashami13, Radko16}. The total decay rate is then $\Gamma = \Gamma_\mathrm{rad} + \Gamma_\mathrm{nr}$, and sets the excited state lifetime $\tau = 1/\Gamma$. We then term $\eta_i$ the intrinsic quantum efficiency, defined as $\eta_i = \Gamma_\mathrm{rad}/(\Gamma_\mathrm{rad} + \Gamma_\mathrm{nr}$). This important figure of merit cannot be extracted from the fluorescence decay alone, since it requires to disentangle the two types of process. To this end, we follow the standard approach that consists of modifying the LDOS in a controlled fashion~\cite{Nikolai19, Drexhage70,Tews70, Kunz80, Brokmann04, Mohtashami13, Radko16}. The subsequent changes in the emitter photophysical properties yield informations about its internal dynamics as well as its coupling to the EM field, given that only the processes associated with $\tau_\mathrm{rad}$ are affected, while the intrinsic non-radiative channels remain unchanged. Indeed, $\Gamma_\mathrm{nr}$ remains constant, while $\Gamma_\mathrm{rad}$ is modified to $\Gamma_P = \Gamma_\mathrm{rad} \rho(\mathbf{r})/\rho_0$, where $\rho_0$ is the LDOS in the infinite, homogeneous medium and $\rho(\mathbf{r})$ is the LDOS in the photonic structure, at a position $\mathbf{r}$. Their ratio is the Purcell factor $F_P(\mathbf{r}) = \rho(\mathbf{r})/\rho_0$. We note that the new EM environment can induce additional non-radiative processes, for instance in the case of plasmonic or other metallic structures where losses can occur in the metal. These losses are included in $\Gamma_P$. They only act on radiative transitions and not on LDOS-independent non-radiative processes, and thus can be treated as photon losses, with no impact on the method. The modified total decay rate reads $\Gamma' = 1/\tau' = \Gamma_P + \Gamma_\mathrm{nr}$. The intrinsic quantum efficiency can then be seen as a scaling factor that tunes the impact of a change in the LDOS on the lifetime~\cite{Nikolai19}, according to $\Gamma' = \Gamma \left( 1 + \eta_i (\rho(\mathbf{r})/\rho_0 - 1 ) \right)$.

\begin{figure}
 \includegraphics[width=\linewidth]{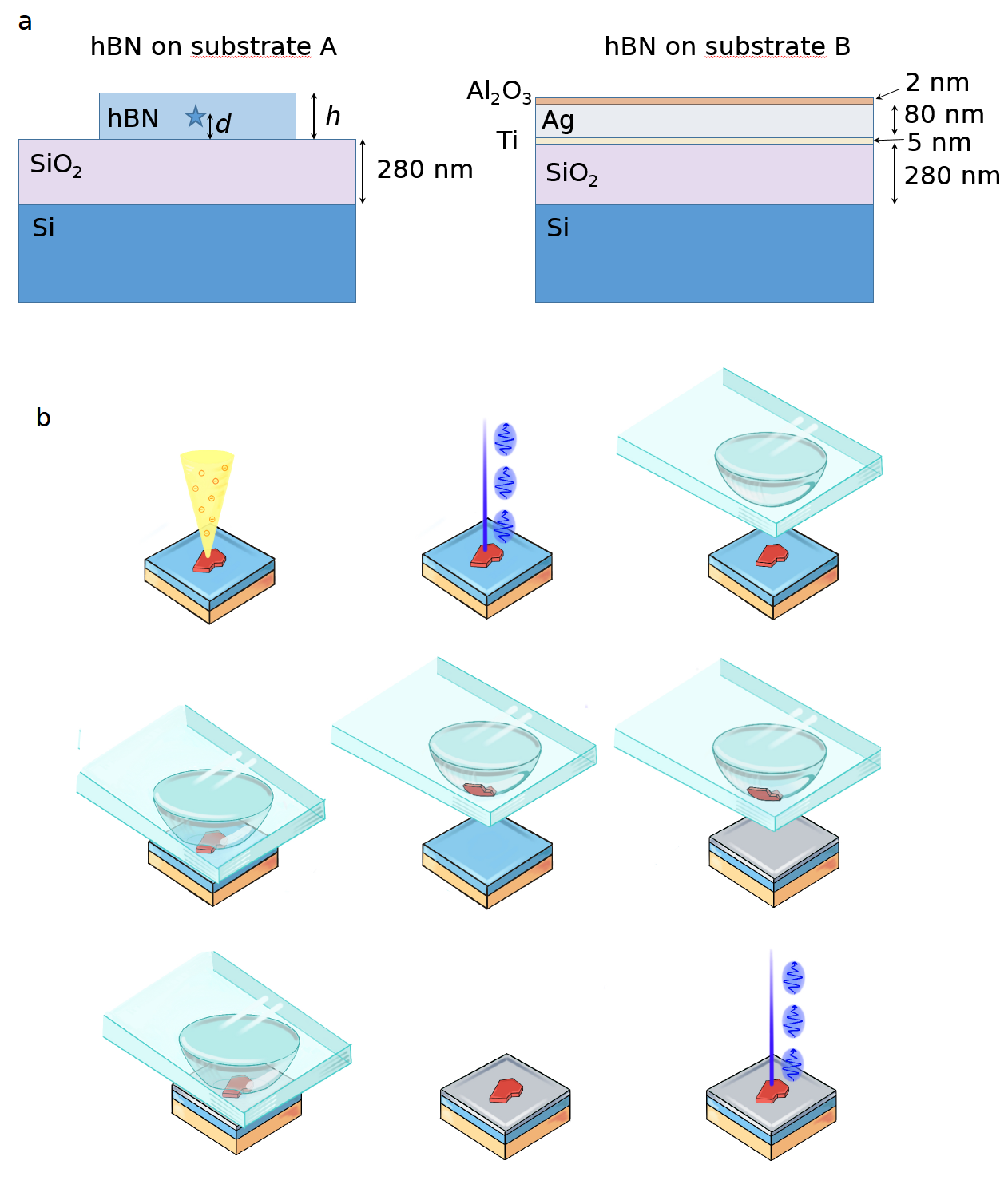}%
 \caption{\label{samples} (a) Scheme of the two structures used in this work. The first substrate (substrate A) is a SiO$_2$/Si sample on which hBN crystals are exfoliated and irradiated. The second substrate (substrate B) is coated with Ag and a top protective layer of Al$_2$O$_3$. (b) Scheme of our experimental process. Exfoliated hBN crystals are exfoliated on substrate~A, irradiated to generate SPEs, characterized in PL, transfered from substrate A to substrate B, and characterized again in PL.}%
 \end{figure}

Experimentally, we perform such comparison by successively integrating the same SPEs in two different structures, as depicted on figure~\ref{samples}. The first structure consists of hBN crystals on a SiO$_2$ (280~nm)/Si substrate (substrate~A on figure~\ref{samples}a). The second is formed by the hBN flake and a bottom 80~nm silver film deposited on SiO$_2$/Si (substrate~B on figure~\ref{samples}a). More details about the sample fabrication are provided in the Methods section. The optical properties of the resulting heterostructures depend on the hBN flake thickness $h$, as well as the vertical position $d$ of the quantum emitters within the flake. The whole protocol for comparing well-identified SPEs in the two different structures is described on figure~\ref{samples}b. We use undoped, high-purity HPHT-grown hBN from NIMS~\cite{Taniguchi07}, which we exfoliate on substrate~A. The chosen flakes exhibit a bright free-exciton peak in cathodoluminescence. This is a signature of low defect density that attributes a high quantum efficiency to the exciton, owing to limited non-radiative recombination processes~\cite{Roux21}. We also select rather thick flakes ($h>70$~nm), which may mitigate the potential surface-induced non-radiative pathways. Five flakes of thicknesses ranging from 70~nm to 220~nm are selected and irradiated using an electron beam of acceleration voltage 3-5~kV to create SPEs~\cite{Fournier21, Gale22}, which are subsequently characterized in photoluminescence (PL). We then transfer the flakes of interest on the Ag substrate (substrate~B) using a dry transfer technique (see Methods) and characterize the sample in PL a second time. Confocal mapping allows identification of the exact same emitters before and after transfer (see Supporting Information).

 \begin{figure*}
 \includegraphics[width=0.95\linewidth]{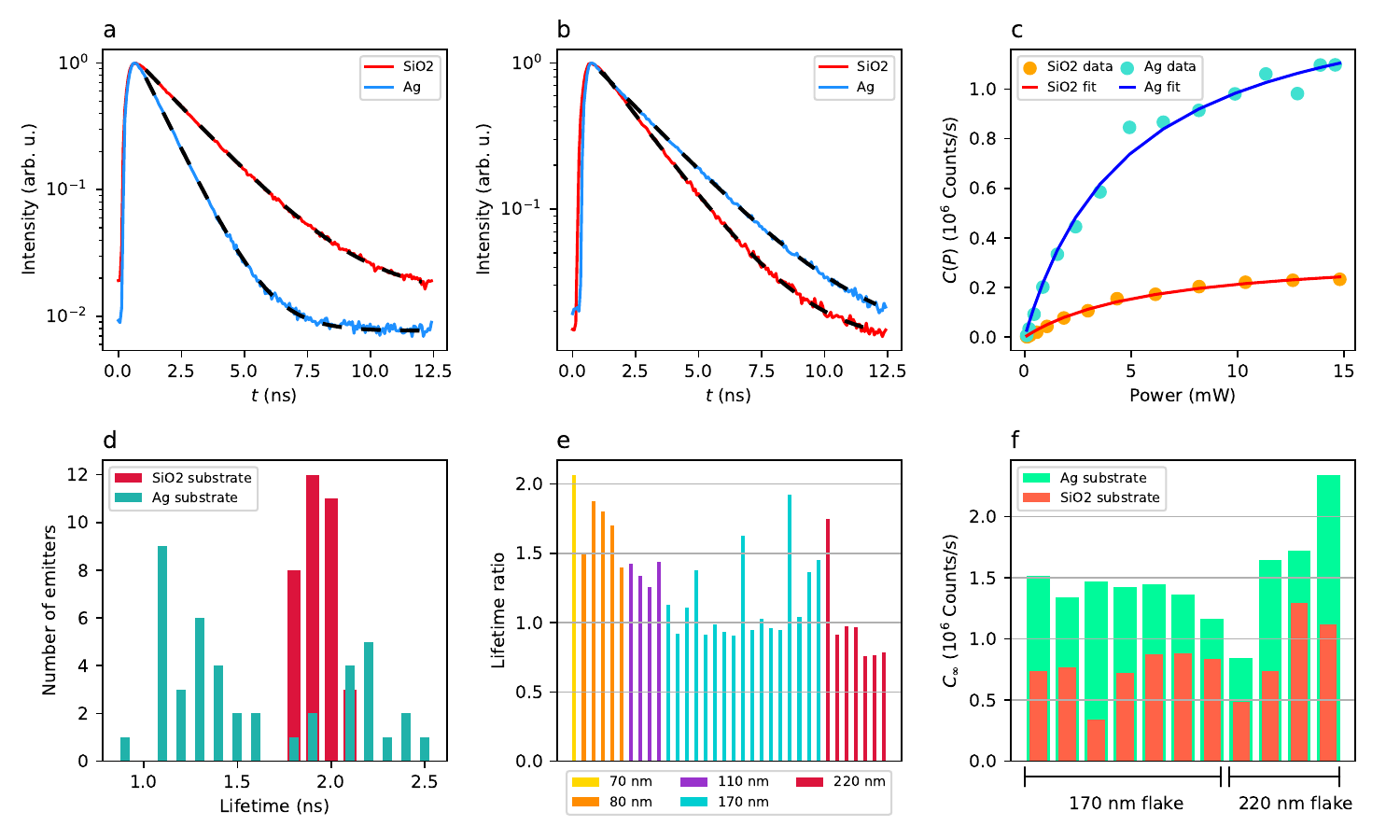}%
 \caption{\label{PL} (a) Fluorescence decay of a SPE on substrate~A (red curve) and B (blue curve), demonstrating lifetime shortening. The dashed black lines are exponential fits to the data. (b) Fluorescence decay of another SPE on substrate~A (red curve) and B (blue curve), demonstrating fluorescence inhibition. The dashed black lines are exponential fits to the data. (c) Saturation curve of a SPE on substrate A (orange dots) and B (light blue dots). Plain curves are fits to the data (see main text). (d) Histogram of the lifetimes measured on substrate~A (red bars) and substrate~B (turquoise bars), showing the modification of the distribution induced by the change of photonic structure. (e) Lifetime ratio $\tau_\mathrm{A}/\tau_\mathrm{B}$ for 34~individual emitters. The bar colors indicate the thickness of the host flake. (f) Saturation count rate $C_\infty$ measured for several emitters on sample~A (orange bars) and sample~B (green bars).}
 \end{figure*}
 
The PL measurements are performed in a confocal microscope at room temperature (see Methods). A 405-nm pulsed laser is used to excite the color centers at a power of 350~$\mu$W. Lifetime and intensity autocorrelation are measured using avalanche photodiodes with a resolution of 65~ps. Additionally, the count rate is measured under cw excitation as a function of the laser power to infer the saturation count rate of the SPEs. This set of measurements is performed on well-identified SPEs before and after transfer from the SiO$_2$ substrate to the Ag substrate. In either case, the emitter lifetime is inferred from the fluorescence decay. The decay of two representative emitters is shown on figure~\ref{PL}a and b, where the blue (red) curves show the measurement before (after) transfer. An exponential fit to the data (shown in dashed black lines) provides the fluorescence lifetime $\tau_\mathrm{A}$ ($\tau_\mathrm{B}$) on substrate~A (B) (see Supporting Information). In the case of the first emitter, the lifetime has decreased from $\tau_\mathrm{A} = $2.06~ns to $\tau_\mathrm{B} = $1.07~ns, indicating lifetime shortening, which is attributed to the Purcell effect. For the second emitter, the lifetime is longer after transfer, increasing from $\tau_\mathrm{A} = $1.88~ns to $\tau_\mathrm{B} = $2.44~ns, thereby demonstrating Purcell inhibition of the emitter fluorescence. The statistics for $\tau_\mathrm{A}$ of all emitters on the SiO$_2$ substrate is shown on figure~\ref{PL}d (red bars). The distribution is narrow and centered around 1.93~ns, with a standard deviation of 0.08~ns, consistently with prior work~\cite{Fournier21}. After transfer on the Ag substrate, the lifetime of most emitters is modified, resulting in the much broader distribution of $\tau_\mathrm{B}$ visible on figure~\ref{PL}d (turquoise bars). The ratio of fluorescence lifetimes $\tau_\mathrm{A}/\tau_\mathrm{B}$ measured for each emitter is plotted on figure~\ref{PL}e. The bar colors identify the thickness of the host flakes. For the three thinnest flakes, the lifetime is always shortened, up to a factor~2. However, in the case of the thickest flakes, the emitter decay is in some cases prolonged on substrate~B with respect to substrate~A. This observation of Purcell inhibition alone demonstrates that the emitter decay is not dominated by non-radiative relaxation, and allows to infer an absolute lower bound for the internal quantum efficiency~$\eta_i$ for the emitters exhibiting fluorescence inhibition. Indeed, $\eta_i = \tau_\mathrm{nr}/(\tau_\mathrm{rad} + \tau_\mathrm{nr})$, where $\tau_\mathrm{rad}$ is the radiative lifetime in the absence of Purcell modification, and $\tau_\mathrm{nr}$ the intrinsic non-radiative lifetime. The excited state lifetime is given by $\tau = (1/\tau_\mathrm{rad} + 1/\tau_\mathrm{nr})^{-1}$. In the hypothetical limit where the emitter is placed in a photonic structure that yields perfect inhibition of the radiative decay (\textit{i.e.} vanishing Purcell factor), the modified lifetime is given by $\tau' = \tau_\mathrm{nr}$, which corresponds to a ratio $\tau/\tau' = \tau/\tau_\mathrm{nr} =  1- \eta_i$. For finite Purcell inhibition, we then have in general $ \tau/\tau' > \tau/\tau_\mathrm{nr} = 1 - \eta_i $, which yields in our case $\eta_i > 0.24$. In the next section, we perform numerical simulations to quantify the fluorescence enhancement and inhibition in our structures, allowing to provide a more accurate estimation of~$\eta_i$. 

We also perform saturation measurements of multiple emitters on both substrates, by recording the count rate as a function of the laser power in cw regime. Figure~\ref{PL}c provides an example of such measurement, for the same emitter, either on substrate~A (orange dots) or B (light blue dots). In this particular case, the change of substrate yielded a clear enhancement of the collection efficiency. The data are fitted using the power dependence of the count rate from a two-level system, $C(P) = C_\mathrm{\infty} / (1 + P_\mathrm{sat}/P)$, allowing to extract the high-power asymptote $C_\mathrm{\infty}$. The saturation power $P_\mathrm{sat}$ is modified from sample~A to sample~B by the Purcell effect as well as by the modified propagation of the laser light due to the photonic structure, which affects the laser intensity at the emitter position due to interference effects between the incident and the reflected laser beam. The asymptote $C_\mathrm{\infty}$ also varies from sample A to~B due to its dependence on both the radiative lifetime and the collection efficiency $\eta_\mathrm{coll}$ as $C_\infty \propto \eta_\mathrm{coll}/\tau_\mathrm{rad}$. However, since it corresponds to a maximum population of the excited state, it is independent of the laser power needed to reach it, and thus of the laser-emitter coupling, allowing to simulate the change of $C_\infty$ without having to account for the modification of $P_\mathrm{sat}$. Figure~\ref{PL}f presents $C_\mathrm{\infty}$ for several emitters, before and after transfer. The enhancement of the collection efficiency by the Ag structure varies depending on the emitter, which can be attributed to the variety of flake thicknesses and vertical positions of the measured emitters, as we confirm in the following section devoted to numerical simulations.

 \begin{figure*}
 \includegraphics[width=\linewidth]{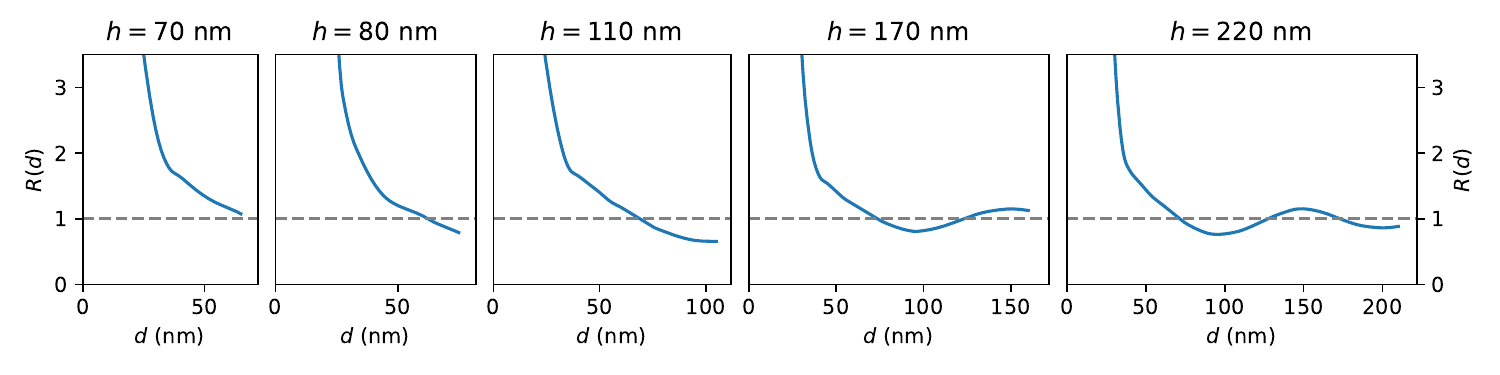}%
 \caption{\label{Purcell} Calculated ratio $R(d)$ of the Purcell factor in structures A and B as a function of the emitter position $d$ for the five considered thicknesses.}%
 \end{figure*}

\section*{Numerical simulations}

To gain quantitative insight into these experimental results, we perform numerical simulations using the finite difference time domain (FDTD) method. We use a point dipole source with in-plane orientation, given that the B-centers have a single dipole oriented in the basal plane~\cite{Fournier21}. The point dipole is located at a distance $d$ from the bottom of the flake (see figure~\ref{samples}). To closely match our experimental protocol, for each vertical position $d$ of the emitter, we calculate the ratio $R(d)$ between the Purcell factor in both structures: $R(d) = F_\mathrm{P}^B(d)/F_\mathrm{P}^A(d)$. Experimentally, this ratio corresponds to the ratio between lifetimes $\tau_\mathrm{A}$ and $\tau_\mathrm{B}$ measured for a given emitter located at a given height $d$. Figure~\ref{Purcell} shows $R(d)$ calculated for five flake thicknesses matching the experimentally investigated samples. In all cases, the Purcell factor diverges when $d$ vanishes, due to the proximity to the metal layer. For the three thinnest flakes, $R(d)$ has a monotonic dependence on $d$, allowing an unequivocal identification of the emitter position from the measured lifetimes $\tau_\mathrm{A}$ and $\tau_\mathrm{B}$. In the case of the two thickest flakes in turn, $R(d)$ oscillates for $d \gtrsim 50$~nm, and in some regions lies below~1, indicating a slight inhibition of the spontaneous emission. These oscillations originate from the varying coupling to the in-plane guided mode, as will be evidenced in the following.

For the 220~nm flake, we calculate the lowest value of $R(d)$ to be 0.76. This value closely matches the highest inhibition of the spontaneous emission measured for a SPE in this flake ($\tau_\mathrm{A}/\tau_\mathrm{B} = 0.765 \pm 0.02$, see figure~\ref{PL}e). Given that our simulations model a purely radiative dipole ($\eta_i = 1$), this excellent agreement shows that the change of lifetime is fully accounted for by the modification of the spontaneous emission rate by the photonic structure and thus demonstrates that $\eta_i = 1.00 \pm 0.02$. To estimate the statistics of $\eta_i$ for the whole ensemble of emitters, we recall that in sample~A (out of the cavity), the lifetime distribution is $\tau = 1.93 \pm 0.08$~ns. Since the statistical spread of the lifetime distribution can originate from both radiative and non-radiative processes~\cite{Mohtashami13}, we can estimate a higher bound of the spread in $\eta_i$ by considering the "worst-case" scenario where the spread in $\tau$ entirely originates from a spread in $\tau_\mathrm{nr}$. In the limit of high quantum efficiency, the standard deviation of $\eta_i$ simply writes $\Delta \eta_i \approx \Delta\Gamma_\mathrm{nr} / \Gamma$, which provides $\eta_i = 0.96 \pm 0.04$ as a conservative estimation of the QE variability for the ensemble of emitters characterized in this work. Based on this argument, in the following we assume $\eta_i \approx 1$ for all emitters.

%

 \begin{figure*}
 \includegraphics[width=\linewidth]{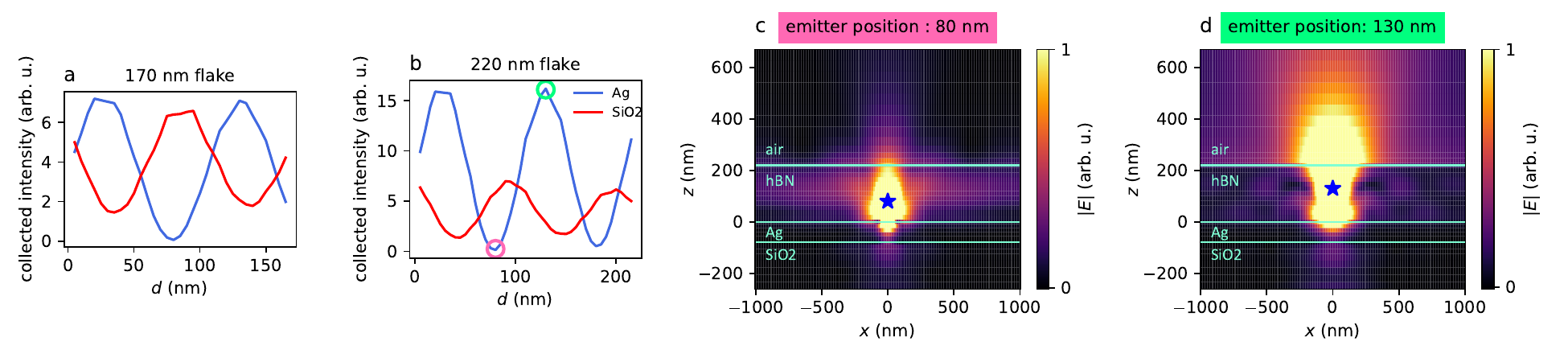}%
 \caption{\label{collection} (a) and (b): Simulated collected intensity from an emitter at position $d$ in the 170~nm flake (a) and in the 220~nm flake (b), on substrate A (red curves) and B (blue curves). (c) and (d): simulated electric field profile for the two emitter positions depicted by the circles on figure~\ref{collection}b.}%
 \end{figure*}

In the case of the 170~nm and 220~nm flakes, the Purcell factor alone does not allow a definite identification of the emitter position, since it is non-monotonic. Therefore, for the two thickest flakes, we also perform FDTD simulations to calculate the power radiated into a top objective of NA = 0.95, matching our experimental setup. The simulations account for the Purcell effect, the modification of the radiation pattern, as well as losses in the metallic layers (see Methods). Therefore, the simulated collected power is proportional to the count rate collected by a two-level emitter driven at saturation. Figure~\ref{collection} shows the calculation results for the two crystals in configurations A and B. The collected intensity oscillates as a function of $d$ in all cases, due to a competition between in-plane and out-of-plane emission. This phenomenon is at the origin of the brightness disparities of SPEs in thick flakes observed in prior work~\cite{Roux22}, and has already been taken benefit of in waveguide structures~\cite{Gerard23}. Interestingly, the position of the nodes and antinodes is different on substrates A and B due to different boundary conditions. This leads to a position-dependent enhancement of the collection efficiency upon transfer from substrate A to B. The variation of the radiation pattern with $d$ is illustrated by the electric field profile shown on figure~\ref{collection}c and d, respectively at a minimum and at a maximum of the collection. In the former case, most of the light is emitted in the plane, while in the latter case, the light is directed to the top with a relatively low divergence (about 40$^{\circ}$ emission angle), which could be efficiently collected by an air objective of NA~$\sim 0.65$. This shows that the metal-dielectric structure constitutes a simple yet effective way to improve the photon collection. Simulations predict collection enhancement up to a factor~8, consistent with our experimental observations. Additionally, the oscillations of the collection enhancement as a function of the emitter position provides an additional information likely to lift the remaining uncertainty about the emitter positioning as inferred only by the Purcell factor.

\section*{Determination of emitter positions}

In this section, we use a maximum likelihood estimation method to infer the emitter positions from the Purcell enhancement and, when necessary, the collection enhancement. We attribute to a measured physical quantity $x_i$ a Gaussian probability distribution function (PDF) $L_{x_i}(d) = f_i(\hat{x}_i|d) = 1/(\sigma_{x_i} \sqrt{2 \pi}) \exp{[- (\hat{x_i} - x(d))/2 \sigma_{x_i}^2]}$, where $x(d)$ is the theoretical (calculated) value of the parameter $x_i$ for a position $d$, $\hat{x_i}$ the measured value, and $\sigma_{x_i}$ accounts for the experimental uncertainties. The likelihood function $L(d)$ is defined as the joint PDF of all parameters $x_i$, \textit{i.e.} $L(d) = \prod_i f_i(\hat{x}_i|d)$. Its maximum provides the most likely emitter position given the measured values of the physical quantities $x_i$, and its width gives the associated confidence interval.

\begin{figure}
 \includegraphics[width=0.8\linewidth]{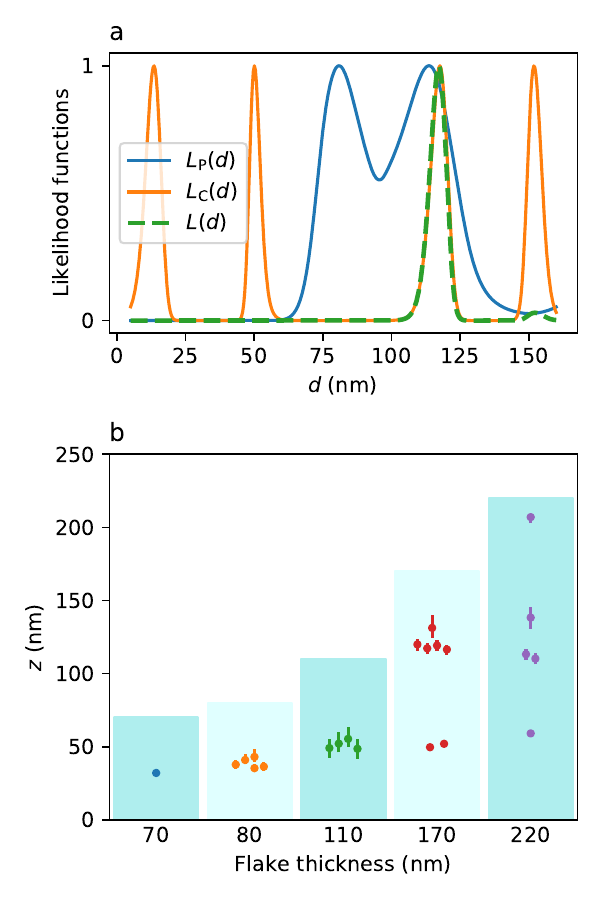}%
 \caption{\label{max_L} (a) Example of likelihood functions for an emitter in the 170~nm flake. Plain curves: partial likelihood functions $L_\mathrm{P}(d)$ and $L_\mathrm{C}(d)$. Dashed green line: joint (total) likelihood function $L(d)$. (b) Maximum likelihood positions of SPEs in the five flakes. The error bars represent the FWHM of the maximum likelihood peak. The dots are horizontally offset for clarity. The blue rectangles symbolize the hBN material for each thickness.}%
 \end{figure}

For the three thin flakes, we use a single parameter $x_i$ which is the Purcell enhancement $\tau_A/\tau_B$, associated with the theoretical value $R(d)$. As expected, the associated likelihood function has in all cases a single maximum. This is however not the case for the 170~nm and 220~nm flakes due to the non-monotonicity of $R(d)$ as visible on figure~\ref{Purcell}. For these two cases, we therefore use two parameters, namely the Purcell enhancement and the collection enhancement. Experimentally, the latter is evaluated through the ratio $C_\infty^\mathrm{B}/C_\infty^\mathrm{A}$, where $C_\infty^{\mathrm{A}(\mathrm{B})}$ is the high power asymptote of the count rate measured in structure A (B). To illustrate the method, we focus on the example of a particular emitter of the 170~nm flake, whose characterization yields a Purcell factor of $0.91$ and a collection enhancement of $1.75$. Figure~\ref{max_L}a shows the single-parameter likelihood functions $L_\mathrm{P}(d)$ and $L_\mathrm{C}(d)$ associated with the Purcell enhancement and the collection enhancement, respectively. For this representative emitter, the Purcell enhancement is associated with two most likely positions (80~nm and 113~nm), as visible through the shape of $L_\mathrm{P}(d)$ (blue curve) and consistently with the simulation shown on figure~\ref{Purcell}. In turn, the collection efficiency is compatible with four positions, as evidenced by $L_\mathrm{C}(d)$ (orange curve). As a result, the (total) likelihood function $L(d) = L_\mathrm{P}(d) L_\mathrm{C} (d)$ has a single maximum, showing that cross-referencing these two observables has allowed to decide the most likely position, resulting in improved accuracy.

The results of the position estimation on the five samples are shown on figure~\ref{max_L}b, where the dots represent the emitter vertical position, the error bars provide the associated confidence level, and the blue rectangle delimits the flakes in the vertical direction. In the three thin flakes, the emitters are found near the center of the flake. In the 170~nm flake, the emitters are also found rather far from the interfaces, and are grouped into two regions, with no emitter found around $d = 90$~nm. As can be seen on figure~\ref{collection}a, this region corresponds to vanishing collection from the top, with predominant photon emission in the plane. Therefore, it cannot be concluded that there are no emitters with $d \approx 90$~nm, since their emission would anyway not be efficiently detected by our setup. Similar considerations can be drawn for the thickest flake, which has two regions of vanishing upwards emission around $d = 80$~nm and $d = 180$~nm, where no emitters are found. The main general observation from the obtained distributions is that most emitters are found more than 30~nm away from the flake boundaries, at the exception of a single emitter on the 220~nm flake. We note that, unusually, the likelihood function of this particular emitter is not single-peaked and has a secondary maximum at position $d = 165$~nm. We therefore cannot strictly exclude that this emitter is actually further away from the top interface.

A possible reason for obtaining estimated positions always far from the interfaces is that the SPEs might be unstable when created too close to the top surface. Indeed, the argument of a vanishing collection efficiency does not apply to SPEs near the interfaces, as clear from figure~\ref{collection}a and~b, which shows sizable signal from dipoles located less than 30~nm away from the flake limits. This observation is consistent with the previously observed difficulty of creating stable B-centers in thin flakes~\cite{Nedic24}, as well as the reported instability of other types of hBN quantum emitters, which exhibit blinking and bleaching for small thicknesses down to monolayers~\cite{Tran16, Stern19}. However, for emitters near the underside, quenching effects due to the proximity of the Ag layer become significant~\cite{Anger06,Vion10}, so that no definitive conclusions concerning their absence or instability can be drawn from our experiment. To stabilize the emitters near the flake surface, potential mitigation strategies could include encapsulation in undoped hBN layers or other suitable materials.

\section*{Conclusion}

We have compared the emission properties of individual hBN color centers in two different planar photonic structures. We took benefit from dry transfer techniques, which allowed us to track individual emitters and study the influence of varying electromagnetic environment on their dynamics. Based on Purcell modification of their lifetime, we have estimated that their quantum efficiency is about unity. We have also inferred the distribution of their vertical positions -- which is a subset of the possible locations of B-centers upon electron beam irradiation. We conclude that they are mostly located away from the flake interfaces, potentially due to a lower stability of SPEs close to the crystal boundaries. Finally, our work demonstrates a simple approach to boost the collection efficiency of quantum emitters in 2D materials, which does not involve demanding nanofabrication techniques.

During the reviewing process, we became aware of a complementary study that confirms the high quantum efficiency of B centers~\cite{Yamamura24}. The slight discrepancy observed between the two estimations can have various origins. The main differences lie in the crystal doping level (we use intrinsic, high-purity material, while ref.~\cite{Yamamura24} uses carbon-doped crystals), and the flake thickness (in our case, the use of thicker flakes could mitigate potential surface-related non-radiative pathways). Besides, some sample preparation steps, in particular electron irradiation, could also play a role in the impact of non-radiative relaxation channels. Overall, it is remarkable that the quantum efficiency of B centers retains high values in a wide range of conditions.\\

\section*{Methods}
\subsection*{Silver mirror fabrication}
The metallic mirror is realized on a SiO$_2$/Si substrate by Ag RF sputtering (Leybold Z400) on a 5~nm thick Ti adhesion layer, followed by depostion of 0.8~nm aluminum. The growth rates are 10~nm/min and 3.5~nm/min for Ag and Al, respectively. The Al layer naturally oxidizes when exposed to ambient atmosphere, thereby realizing a $\sim 2$~nm Al$_2$O$_3$ protective layer for the underneath Ag. XPS measurements ensure that all Al is oxidized, and that it covers hermetically the underlying Ag layer.

\subsection*{Exfoliation and transfer}
Exfoliation is performed on substrate~A from bulk crystals using commercial silicone elastomer films (Gel-Pak). The flake thicknesses are measured by atomic force microscopy. Transfer to substrate~B is done using dry transfer~\cite{Wang13} based on a glass slide with a home-made hemispherical polydimethylsiloxane stamp coated with polypropylene carbonate. The picking step from substrate~A is done at 25 to 30~$^{\circ}$C. The transfer is done by approaching the stamp at 40~$^{\circ}$C and increasing the temperature to 80 to 100~$^{\circ}$C after contact to release the flake on substrate~B. The process ends with a cleaning step (acetone + isopropanol).

\subsection*{Optical measurements}
Photoluminescence measurements are performed in a room-temperature confocal microscope using a NA$ = 0.95$ air objective. The emitter are excited using a PicoQuant laser diode emitting at 405~nm, and working in either pulsed or cw regime. The photoluminescence signal is filtered using a bandpass fluorescence filter centered at 442~nm, and collected by two avalanche photodiodes in Hanbury Brown and Twiss configuration. Intensity autocorrelation is measured for all SPEs to ensure that we only consider individual emitters (\textit{i.e.} $g^{(2)}(0) < 0.5$).

\subsection*{Numerical simulations}
Numerical simulations are performed using Lumerical software, in 3D. The SPE is modeled by a point dipole with in-plane orientation. The complex refractive index of the Ag layer is measured by ellipsometry.

\section*{Acknowledgments}

This work is supported by the French Agence Nationale de la Recherche (ANR) under reference ANR-21-CE47-0004-01 (E$-$SCAPE project). This work has also been supported by Region \^Ile-de-France in the framework of DIM QuanTiP. The Authors acknowledge K. Watanabe and T. Taniguchi for providing hBN crystals, as well as J.~Barjon, C.~Arnold and G. Colas des Francs for many useful discussions.

\end{document}